\documentclass{article}
\usepackage{emulateapj}
\usepackage{apjfonts}

\begin{document}

\submitted{To appear in The Astrophysical Journal Letters}

\title{Photospheric Abundances of the Hot Stars in NGC~1399 and 
Limits on the Fornax Cluster Cooling Flow}


\author{Thomas M. Brown,  Henry C. Ferguson}

\affil{Space Telescope Science Institute, 3700 San Martin Drive,
Baltimore, MD 21218.  tbrown@stsci.edu, ferguson@stsci.edu}

\author{Robert W. O'Connell}

\affil{Department of Astronomy, University of Virginia, P.O. Box 3818,
Charlottesville, VA 22903.  rwo@virginia.edu}

\author{Raymond G. Ohl}

\affil{Code 551, NASA Goddard Space Flight Center, Greenbelt, MD 20771. 
rohl@pop500.gsfc.nasa.gov}

\begin{abstract}

We present far-UV spectroscopy of the giant elliptical galaxy
NGC~1399, obtained with the Far Ultraviolet Spectroscopic Explorer.
Of all quiescent ellipticals, NGC~1399 has the strongest known ``UV
upturn'' -- a sharp spectral rise shortward of 2500~\AA.  It is now
well-established that this emission comes from hot horizontal branch
(HB) stars and their progeny; however, the chemical composition of
these stars has been the subject of a long-standing debate.  For the
first time in observations of any elliptical galaxy, our spectra
clearly show photospheric metallic absorption lines within the UV
upturn.  The abundance of N is at 45\% solar, Si is at 13\% solar, and
C is at 2\% solar.  Such abundance anomalies are a natural
consequence of gravitational diffusion.  These photospheric abundances
fall in the range observed for subdwarf B stars of the Galactic field.

Although NGC~1399 is at the center of the Fornax cluster, we find no
evidence for \ion{O}{6} cooling flow emission.  The upper limit to
$\lambda\lambda 1032,1038$ emission is $3.9 \times 10^{-15}$ erg
s$^{-1}$ cm$^{-2}$, equivalent to 0.14~$M_{\odot}$ yr$^{-1}$, and less 
than that predicted by simple cooling flow models of the NGC~1399 X-ray
luminosity.

\end{abstract}

\keywords{galaxies: abundances -- galaxies: elliptical -- 
galaxies: stellar content -- ultraviolet: galaxies}

\section{INTRODUCTION}

The spectra of elliptical galaxies and spiral galaxy bulges exhibit a
strong upturn shortward of 2700~\AA, dubbed the ``UV upturn.''  The
phenomenon was among the first major discoveries in UV extragalactic
astronomy (Code 1969), and the implied existence of a hot stellar
component in ellipticals contradicted the traditional picture of
early-type galaxies as old, cool, passively-evolving populations.
Early debate covered many possible candidates for the UV upturn origin
(see O'Connell 1999), but today we know from UV spectroscopy (Ferguson
et al.\ 1991; Brown et al.\ 1997) and imaging (Brown et al.\ 2000)
that this UV emission comes from a minority population of extreme
horizontal branch (EHB) stars and their progeny.  These stars appear
to be analogs of the subdwarf B (sdB) and O (sdO) stars of the local
Galactic field, but the populations may differ considerably in the two
cases.

The UV upturn is not completely understood, however.  Long before the
observational proof of an EHB origin, EHB stars were a
strong theoretical candidate, and there have been two schools of
thought regarding their nature.  One possibility is that the EHB stars
reside in the metal-poor tail of a wide metallicity distribution
(e.g., Park \& Lee 1997).  Another possibility is that the EHB stars
lie at high metallicity, reflecting the composition found in the
cooler, dominant populations of elliptical galaxies (e.g., Brocato et
al.\ 1990; Bressan, Chiosi, \& Fagotto 1994; Greggio \& Renzini 1990;
Dorman, O'Connell, \& Rood 1995; Horch, Demarque, \& Pinsonneault
1992; Brown et al.\ 1997).  The metallicity controversy is tied to the
``second parameter'' debate, which seeks to understand the role of
parameters (besides metallicity) that govern HB morphology in globular
clusters; possible candidates include
age, mass loss, He abundance, rotation, and cluster dynamics (see Fusi
Pecci \& Bellazzini 1997).

Characterized by the $m_{1550}-V$ color, the UV upturn shows
surprisingly strong variation (ranging from 2.05--4.50 mag) in nearby
quiescent early-type galaxies (Bertola, Capaccioli, \& Oke 1982;
Burstein et al.\ 1988), even though the spectra of ellipticals at
longer wavelengths are qualitatively very similar.  The $m_{1550}-V$
color is positively correlated with the strength of Mg$_2$ line
absorption in the $V$ band (i.e., bluer at higher line strengths),
opposite to the behavior of optical color indices (Burstein et al.\
1988).  Both sides in the EHB metallicity debate see this correlation
as further proof of their own scenario.  The metal-poor school reasons
that galaxies with strong UV emission are more massive and formed
earlier; such galaxies have higher {\it mean} metallicity, as tracked
through optical indices, even though the upturn comes from stars in
the metal-poor tail.  In this scenario, age is the dominant ``second
parameter'' driving HB morphology, with the metal-poor HB becoming
more blue with age, thus producing more UV emission in massive old
galaxies.  In contrast, the metal-rich school argues that the trend
for the HB distribution to become redder at increasing metallicity can
be reversed at high metallicity, due to increased He abundance and
possibly a higher mass-loss rate on the red giant branch (RGB), both
of which lead to more metal-rich EHB stars and a subsequent increase
in UV emission from the population.  These hypotheses imply
different ages for the stellar populations in these galaxies.  Ages
exceeding those of Galactic globular clusters are required under the
metal-poor Park \& Lee (1997) hypothesis, while ages as low as 8~Gyr
are allowed in the metal-rich Bressan et al.\ (1994) model.

With this debate in mind, we obtained observations of NGC~1399 with
the Far Ultraviolet Spectroscopic Explorer (FUSE; Moos et al.\ 2000).
NGC~1399 is the giant elliptical galaxy at the center of the Fornax
cluster.  It has the strongest UV upturn of any quiescent elliptical
galaxy measured to date ($m_{1550}-V = 2.05$ mag; Burstein et al.\
1988).  Although the UV upturn correlates well with optical
metallicity indices, such indices track the composition of the cool
population (RGB and main sequence stars).  Our far-UV spectroscopy is
intended to measure the photospheric abundances in the hot stars
producing the UV upturn -- measurements not possible in earlier
spectroscopy of elliptical galaxies (Brown et al.\ 1997; Ferguson et
al.\ 1991), due to the low signal-to-noise and resolution.  In
addition, X-ray observations suggest NGC~1399 is the host to a
significant cooling flow (Bertin \& Toniazzo 1995; Rangarajan et al.\
1995), making it an interesting target for \ion{O}{6} emission
measurements.  We present here an analysis of our FUSE data for
NGC~1399; future papers will present our observations of other
ellipticals, which are not yet complete.

\section{OBSERVATIONS}

We observed NGC~1399 on 12--13 Dec 2000 and 3 Oct 2001 for 28 ksec,
with 20 ksec during orbital night.  The spectroscopic resolution is
limited by the velocity dispersion of the galaxy (331 km s$^{-1}$;
Burstein et al.\ 1988), so we scheduled observations in the default
$30\arcsec \times 30\arcsec$ low-resolution square aperture.  The data
were obtained in time-tag mode, recording the arrival time and
detector position of each photon.

We reprocessed the data using the CALFUSE software, version 2.0.5.
Scattered light and airglow emission lines both increase during orbital
day, so we rejected daytime data.  We also implemented pulse-height
screening appropriate for very faint targets, by raising the minimum
pulse height in CALFUSE from 0 to 4, thus discarding more of the
detector dark counts with little loss of source counts.

Useful data were obtained in all detector segments, each with its own
wavelength bins and instrumental resolution.  Because the velocity
dispersion limits the useful resolution, we coadded all of
the data onto a linear wavelength scale (0.25~\AA\ bins), with
propagation of statistical errors.  The bins in our coadded data are
much larger than the nominal FUSE wavelength bins ($\sim 0.01$~\AA),
but they adequately sample the resolution elements in the spectrum ($>
1$~\AA\ for velocity-broadened features). We masked bad pixels when
coadding the data, because CALFUSE does not account for small-scale
flat-field features.  Note that the wavelength solution in this
version of CALFUSE is improved significantly, although these
corrections are unimportant in our large wavelength bins.
Furthermore, although fixed pattern noise can be an issue in
high-resolution FUSE data, it is negligible for our large bins and
coadded detector segments.

\section{STELLAR MODELS}

Our FUSE spectra offer the first clear detection of metallic absorption
lines from the hot stars responsible for the UV upturn.
Previously, Brown et al.\ (1997) fit composite models to
the spectra of elliptical galaxies observed with the Hopkins
Ultraviolet Telescope (HUT), including NGC~1399.  Those spectra were
obtained through large apertures, with coverage from
900--1800~\AA, but the low signal-to-noise and resolution hampered
detection of individual absorption lines from elements heavier than H.
Brown et al.\ (1997) integrated synthetic spectra over EHB
evolutionary tracks from Dorman, Rood, \& O'Connell (1993), using
synthetic spectra appropriate for the HUT resolution (Brown,
Ferguson, \& Davidsen 1996).  The models reproduce well the entire far-UV
spectral energy distribution, and show that the light from elliptical
galaxies is dominated by a population of EHB stars and their post-HB
progeny, the AGB-Manqu$\acute{\rm e}$ stars (Brown et al.\ 1997).

To analyze the FUSE spectra of NGC~1399, we integrated the
best-fitting EHB model from Brown et al.\ (1997), using new
high-resolution synthetic spectra calculated at representative
points along the EHB evolutionary path.  The track assumes a small HB
envelope mass (0.016 $M_\odot$), high metallicity ([Fe/H]=0.71), and
high main-sequence He abundance ($Y = 0.45$).  Note that approximately
half of the far-UV flux in this track comes from the EHB itself, with
the rest coming from the AGB-Manqu$\acute{\rm e}$ phase.  
We computed solar abundance atmospheres using the ATLAS9 code (Kurucz
1993), under the assumption of local thermodynamic equilibrium (LTE).
Synthetic spectra were then calculated using the SYNSPEC code (Hubeny,
Lanz, \& Jeffery 1994), again assuming LTE, but with varying
abundances of C, N, Si, Fe, and Ni (holding other elements at solar
abundance).  Note that this code has been used
previously to model FUSE and HUT data of sdB stars (Ohl, Chayer, \&
Moos 2000; Brown et al.\ 1996). We used the Kurucz (1993) line list in
our spectral synthesis, including lines predicted from atomic physics
but not yet measured in the laboratory.  Weak lines that are
experimentally unconfirmed are potentially prone to position and
wavelength errors, but they provide the necessary pseudocontinuum.

No Galactic foreground reddening is applied to the models.  Both the
maps of Burstein \& Heiles (1984) and Schlegel, Finkbeiner, \& Davis
(1998) suggest that $E(B-V)$ is nearly 0~mag along a line of sight
toward NGC~1399.

To compare the composite models to the FUSE data, we defined spectral
indices of C, N, and Si (Table 1).  Each index measures the ratio of
the mean flux in a line region to that in continuum regions, selected
to be free of strong features from other elements.  The C and Si
indices come from the ratio in one section of the spectrum, whereas
the N index comes from the average in two sections.

\vskip 0.175in

\noindent
\parbox{3.25in}{
{\sc Table 1:} Composition Indices

\begin{tabular}{lccc}
\tableline
         & Cont. & Line        & Dominant Transitions    \\
Ion      & (\AA)      & (\AA)       & (\AA)    \\
\tableline                                                        
\ion{C}{3} & 1170--1173   & 1174--1177  & 1174.93 1175.26 1175.59  \\
           & 1177--1180 &             & 1175.71 1175.99 1176.37  \\
\ion{N}{3} & 986--988 & 989--993    & 989.80 991.51 991.58     \\
           & 994--996 &             &                          \\
\ion{N}{2} & 1074--1078 & 1083--1087  & 1083.99 1084.56 1084.58  \\
           & 1092--1096 &             & 1085.53 1085.55 1085.70  \\
\ion{Si}{3} & 1105--1107 & 1107--1114  & 1108.36 1109.94 1109.97 \\
            & 1114--1116 &             & 1113.17 1113.20 1113.23 \\ 
\tableline
\end{tabular}}

\section{RESULTS}

\subsection{Photospheric Abundances}

By varying the abundances in the synthetic spectra, we find that
the composition that best reproduces the line indices
measured in the data is C at $0.02\pm 0.01$ solar, N at $0.45\pm 0.15$
solar, and Si at $0.13 \pm 0.06$ solar (Figure 1).  The errors
represent statistical uncertainties in each line index, transformed to
abundance uncertainties.  This result does not significantly depend
upon our chosen evolutionary track; although the best-fitting track
from Brown et al.\ (1997) has enhanced He and metallicity, our derived
abundances remain unchanged (within 1$\sigma$) if we use other
unenhanced tracks that reproduce the far-UV spectrum.

Fe and Ni lines blanket the entire FUSE region, but \linebreak

\hskip 0.25in \vskip -0.2in
\parbox{6.5in}{\epsfxsize=6.5in \epsfbox{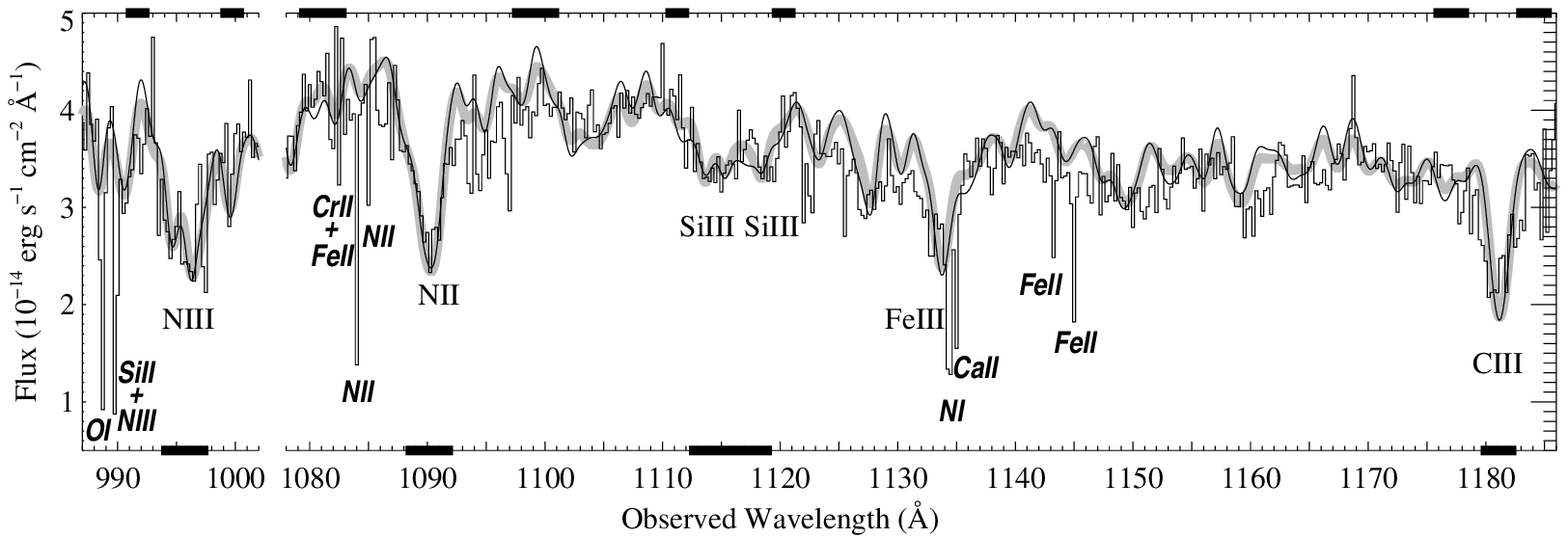}}

\hskip 0.05in
\parbox{6.7in}{\small {\sc Fig.~1--}
Sections of the FUSE NGC~1399 spectrum (black
histogram). Photospheric (roman) and Galactic ISM (italics) features
are easily distinguished by their widths.  The composite model is
shown with photospheric Fe and Ni at solar (thick grey curve) and 5
times solar (thin curve) abundance.  Thick black lines at the top and
bottom of the plot denote the regions used for spectral indices (Table
1).  The model normalization changes by 10\% at the break (the
absolute flux calibration varies by $\sim$10\% between detector
segments).  }

\vskip 0.2in

\noindent
do not
significantly affect the derived abundances of the light elements.  To
demonstrate the systematic errors from Fe and Ni assumptions, we
varied the abundances of Fe and Ni over a extremely large range: from
0.01 solar to 5.0 solar.  Throughout this range, the derived
abundances of C, N, and Si remain within their 1$\sigma$ statistical
uncertainties.  We show the same evolutionary model, when Fe and Ni
are enhanced, in Figure 1; the increased Fe and Ni severely depressed
the continuum, so the curve has been renormalized upward.  Note
that diffusion processes tend to deplete the photospheric abundances
of the light elements in sdB stars (Lamontagne et al.\ 1985;
Lamontagne, Wesemael, \& Fontaine 1987), but Fe can be radiatively
supported; e.g., Fe enhancement is thought to be the mechanism for the
pulsating sdB stars (Charpinet et al.\ 1997), which show no
photospheric Fe depletion (Heber, Reid, \& Werner 2000).

The C and Si line indices rely upon lines that are not resonance
transitions, and thus should not be contaminated from any interstellar
medium (ISM) in NGC~1399 -- i.e., the features in the FUSE spectrum
should be completely photospheric.  In contrast, the N index comes from
series of resonance lines.  Because the NGC~1399 spectrum is
redshifted by 1424 km s$^{-1}$, narrow absorption features from these
same transitions in the Galactic ISM can be seen in the FUSE spectrum,
offset by $\sim 5$~\AA\ to the blue.  If NGC~1399 has a significant
cold ISM, the redshifted N features could have both ISM and photospheric
contributions.

However, when looking at purely interstellar features in the FUSE
data, we see no evidence for absorption from a cold ISM in NGC~1399.
The \ion{N}{2} and \ion{N}{3} features in Table 1 can arise from both
hot stellar photospheres and the ISM, but other resonance transitions
will arise only from an ISM.  These can be used to discern how much
contamination of photospheric features we should expect from an ISM in
NGC~1399.  E.g., strong absorption from \ion{N}{1} (1134.2~\AA,
1134.4~\AA, and 1135.0~\AA) and \ion{Ca}{2} (1135.5~\AA\ and
1135.6~\AA) from the Galactic ISM is present in the FUSE spectrum
(narrow lines in Figure 1), but there is no corresponding strong
absorption in the NGC~1399 frame (1140~\AA, with a width of
1.3~\AA). The same can be said of \ion{C}{2} $\lambda 1036$~\AA\ and
\ion{O}{1} $\lambda 1039$~\AA\ (only Galactic features are present).
Thus, the \ion{N}{2} and \ion{N}{3} spectral indices have little, if
any, contribution from an ISM in NGC~1399, and should track
photospheric abundances only. \\

\vspace{2.75in}

\subsection{Cooling Flow Emission}

NGC~1399 lies at the center of the Fornax cluster.  The X-ray
luminosity of the galaxy implies a large cooling flow, with a mass
deposition of at least 2 $M_{\odot}$~yr$^{-1}$ (Bertin \& Toniazzo
1995; Rangarajan et al.\ 1995).  Cooling flow models predict that this
X-ray emission should be accompanied by \ion{O}{6}
$\lambda\lambda$1031.9,1037.6~\AA\ emission, radiated as the hot gas
cools from $\sim 10^7$~K to $\sim 10^5$~K.  Bregman, Miller, \& Irwin
(2001) did not find such emission in their FUSE observations of
NGC~1404 (upper limit 0.3 $M_{\odot}$~yr$^{-1}$), an elliptical galaxy
near the center of the Fornax cluster, but its X-ray luminosity
suggests a smaller cooling flow deposition rate than that for NGC~1399
(Bertin \& Toniazzo 1995). Thus, NGC~1399 offers a better opportunity
for finding the warm gas from the Fornax cooling flow.

We see no \ion{O}{6} emission in our NGC~1399 spectrum (Figure 2);
indeed, the observed flux at the location of the 1032~\AA\ (rest
frame) component is slightly lower than the expected flux from the stellar
continuum.  To set upper limits, we first fit an EHB model with the
abundances determined above.  The fit region was from 1035--1045~\AA,
and we included \ion{C}{2} and \ion{O}{1} absorption lines from our
own Galactic ISM.  Note that these strong Galactic features are not
accompanied by equivalent features in the NGC~1399 frame, again
implying a lack of a cold ISM.  Once the best fit was achieved, these
components were fixed, and then two components representing the
\ion{O}{6} \linebreak

\smallskip

\parbox{3.25in}{\epsfxsize=3.25in \epsfbox{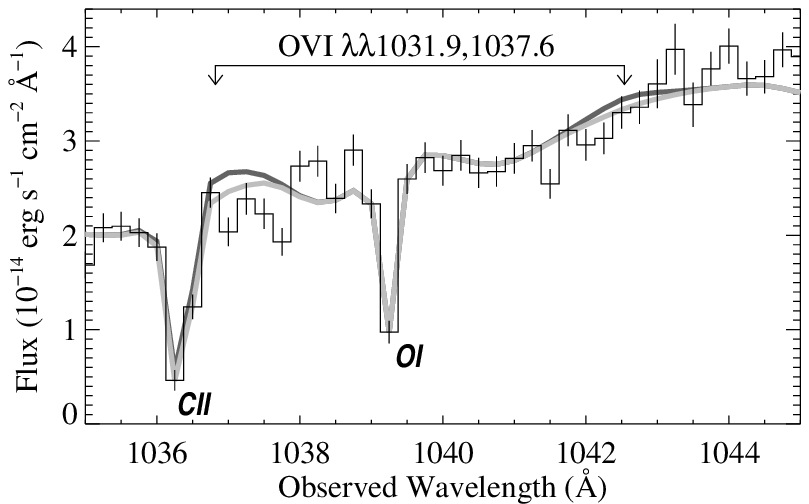}}

\smallskip
\hskip -0.1in
\parbox{3.4in}{\small {\sc Fig.~2--}
The NGC~1399 spectrum
at the expected location of \ion{O}{6} emission (histogram with error bars). 
The curves show the stellar model (light grey), including Galactic
ISM absorption (italic labels), and the
3$\sigma$ upper limit to the redshifted \ion{0}{6} emission (dark grey).}

\noindent
doublet were added.  The wavelengths and widths were fixed
at the NGC~1399 redshift and velocity dispersion, and the flux ratio
was fixed at 2:1 for the 1032 and 1038~\AA\ components.  We find a
3$\sigma$ upper limit of f$_{\lambda 1032} = 2.6 \times 10^{-15}$ erg
s$^{-1}$ cm$^{-2}$ (and thus f$_{\lambda \lambda 1032,1038} =
3.9\times 10^{-15}$ erg s$^{-1}$ cm$^{-2}$ for the total \ion{O}{6}
emission).  As in Bregman et al.\ (2001), we assume that $L_{\lambda
1032} = 0.9 \times 10^{39} \dot{M}$ erg s$^{-1}$ (a compromise between
the relations for isochoric and isobaric cooling).  Thus, the
3$\sigma$ upper limit to the mass deposition rate implied from the
\ion{O}{6} emission is 0.14 $M_{\odot}$ yr$^{-1}$, for the $3\times
3$~kpc region sampled by the FUSE aperture (assuming $m-M = 31.52$
mag; Ferrarese et al.\ 2000).  The limit becomes twice as large if the
fit is restricted to the cleaner region around 1038~\AA.  Models of the
X-ray luminosity imply a larger rate of $\approx 0.4~M_{\odot}$~yr$^{-1}$
within the FUSE aperture (Rangarajan et al.\ 1985).

\section{DISCUSSION}

We have obtained the first clear detection of metallic absorption features
in the UV upturn.  The photospheric composition of the hot population
in NGC~1399 appears to reflect that seen in the field sdB stars of the
Milky Way.  In the nearby Galactic field population, sdB and sdO stars
often show moderate to severe depletions of the light elements,
attributed to diffusion.  Typically, N appears near solar abundance,
while C and Si are depleted (Lamontagne et al.\ 1985; Lamontagne et
al.\ 1987).

Several other processes could skew the photospheric abundances in EHB
stars.  Greggio \& Renzini (1990) noted that hot HB stars of high
metallicity would have very thin H envelopes; these could be
wind-ejected at mass-loss rates of $\sim 10^{-10}~M_{\odot}$
yr$^{-1}$, resulting in surface compositions rich in He and N and poor
in C and O.  CN processing on the RGB is another possible effect
(Sweigart \& Mengel 1979). Assuming initial abundances at the solar
C:N ratio and C+N nuclei conservation, the observed C and N abundances
would imply primordial abundances of 0.12 solar, in the absence of
diffusion.

Galactic field sdB stars tend to show depletion of the light elements
from their original near-solar abundances, but little is known about
the effects of diffusion on metal-poor EHB stars, because EHB stars in
globular clusters are too distant to obtain spectroscopy easily.
Diffusion is the largest uncertainty when associating the photospheric
abundances of EHB stars with their primordial abundances.  The
photospheric abundance of a given element can be affected by both
gravitational settling and radiative levitation, and thus the
photospheric abundance can be either reduced or enhanced relative to
the primordial abundances that drive the evolution of a star.  Some
calculations have shown that abundance enhancements via radiative
levitation can be especially strong in the low-metallicity regime,
because the absorption lines are not saturated (Michaud, Vauclair, \&
Vauclair 1983). 

Taken at face value, our photospheric abundances do not resolve the
metallicity debate for the origin of the UV upturn; they are neither
extremely metal-poor nor extremely metal-rich.  Tying these
photospheric abundances to formation abundances will require further
theoretical and observational work.  In particular, UV observations of
sdB stars in globular clusters are needed to demonstrate how the light
elements are affected by diffusion processes in the metal-poor regime,
showing whether gravitational settling or radiative levitation
dominates.  However, given the tendency for light element abundances
to be diminished by diffusion, it seems unlikely that the UV emission
in NGC~1399 comes from EHB stars with extremely low formation
abundances (e.g., at $\rm -2.4 < [Fe/H] < -1.0$, as suggested by Park \&
Lee 1997); it seems more plausible that these stars started with high
abundances that were then depleted, but at this point such
conclusions are very speculative.

NGC~1399 is thought to be the center of a cooling flow in Fornax.  Our
upper limit to the \ion{O}{6} emission in the FUSE aperture is 
lower than expected from simple cooling flow models of the X-ray emission.
Bregman et al.\ (2001) suggested that the lack of \ion{O}{6} emission
in NGC~1404 could be due to an additional source of heating, but noted
that NGC~1404 has no detected central radio source.  NGC~1399 hosts a
two-lobed radio source (Sadler, Jenkins, \& Kotanyi 1989), which may
help explain the lack of \ion{O}{6} emission.

\acknowledgements

Support for this work was provided by NASA through the FUSE Guest
Investigator program.  The authors gratefully acknowledge support from
NASA grant NAS 5-9696 to the Catholic University of America.  The
authors wish to thank A. Sweigart and W. Landsman for useful insight
and discussions.


\begin{references}

\reference{BT95}
Bertin, G., \& Toniazzo, T. 1995, ApJ, 451, 111

\reference{BCO82}
Bertola, F., Capaccioli, M., \& Oke, J. B. 1982, ApJ, 254, 494

\reference{B01}
Bregman, J.N., Miller, E.D., \& Irwin, J.A. 2001, ApJ, 553, L125

\reference{B94}
Bressan, A., Chiosi, C., \& Fagotto, F. 1994, ApJS, 94, 63

\reference{B90}
Brocato, E., Matteucci, F., Mazzitelli, I., \& Tornambe, A. 1990, 
ApJ, 349, 458

\reference{B00}
Brown, T.M., Bowers, C.W., Kimble, R.A., Sweigart, A.V., \& Ferguson, H.C.
2000, ApJ, 532, 308

\reference{B96}
Brown, T.M., Ferguson, H.C., \& Davidsen, A.F. 1996, ApJ, 472, 327

\reference{B97}
Brown, T.M., Ferguson, H.C., Davidsen, A.F., \& Dorman, B. 1997, ApJ, 482, 685
 
\reference{B88}
Burstein, D., Bertola, F., Buson, L. M., Faber, S. M., \&
Lauer, T. R. 1988, ApJ, 328, 440

\reference{BH84} 
Burstein, D., \& Heiles, C. 1984, ApJS, 54, 33

\reference{C97}
Charpinet, S., Fontaine, G., Brassard, P., Chayer, P., 
Rogers, F.J., Iglesias, C.A., \& Dorman, B. 1997, ApJ, 483, 123

\reference{C69}
Code, A.D. 1969, PASP, 81, 475

\reference{DOR95}
Dorman, B., O'Connell, R. W., \& Rood, R. T. 1995, ApJ, 442, 105

\reference{DRO93}
Dorman, B., Rood, R.T., \& O'Connell, R.W. 1993, ApJ, 419, 596

\reference{F91}
Ferguson, H.C., et al.\ 1991, ApJ, 382, L69

\reference{F00}
Ferrarese, L., et al. 2000, ApJ, 529, 745

\reference{FB97} 
Fusi Pecci, F., \& Bellazzini, M. 1997, in The Third
Conference on Faint Blue Stars, ed. A.G.D. Phillip, J. Liebert, 
R.A. Saffer, \& D.S. Hayes (Schenectady: L. Davis Press), 255

\reference{GR90}
Greggio, L., \& Renzini, A. 1990, ApJ, 364, 35

\reference{HRW00}
Heber, U., Reid, I.N., \& Werner, K. 2000, A\&A, 363, 198

\reference{HDP92}
Horch, E., Demarque, P., \& Pinsonneault, M. 1992, ApJ, 388, L53

\reference{HLJ94}
Hubeny, I., Lanz, T., \& Jeffery, C. S. 1994, in Newsletter on Analysis
of Astronomical Spectra No. 20, ed. C. S. Jeffery (CCP7; St. Andrews: St.
Andrews Univ.), 30

\reference{K93}
Kurucz, R. L. 1993, CD-ROM 13, ATLAS9 Stellar Atmosphere 
Programs and 2 km/s Grid (Cambridge: Smithsonian Astrophys. Obs.)

\reference{L87}
Lamontagne, R., Wesemael, F., \& Fontaine, G. 1987, 318, 844

\reference{L85}
Lamontagne, R., Wesemael, F., Fontaine, G., \& Sion, E.M. 1985, 299, 496

\reference{M83}
Michaud, G., Vauclair, G., \& Vauclair, S. 1983, ApJ, 267, 256

\reference{M00}
Moos, H.W., et al.\ 2000, ApJ, 538, L1

\reference{O99}
O'Connell, R. W. 1999, in ARA\&A, 37, 603

\reference{O00}
Ohl, R.G., Chayer, P., \& Moos, H.W. 2000, ApJ, 538, L95

\reference{PL97}
Park, J.-H., \& Lee, Y.-W. 1997, 476, 28

\reference{R95}
Rangarajan, F.V.N., Fabian, A.C., Forman, W.R., \& Jones, C. 1995, MNRAS,
272, 665

\reference{SJK89}
Sadler, E.M., Jenkins, C.R., \& Kotanyi, C.G. 1989, MNRAS, 240, 591

\reference{SFD98}
Schlegel, D. J., Finkbeiner, D. P., \& Davis, M. 1998, ApJ, 500, 525

\reference{SM79}
Sweigart, A.V., \& Mengel, J.G. 1979, ApJ, 229, 624

\end{references}
\end{document}